# Supersymmetric Gauged Scale Covariance in Ten and Lower Dimensions

Hitoshi NISHINO[1] and Subhash RAJPOOT[2]

*Department of Physics & Astronomy*
*California State University*
*1250 Bellflower Boulevard*
*Long Beach, CA 90840*

## Abstract

We present globally supersymmetric models of gauged scale covariance in ten, six, and four-dimensions. This is an application of a recent similar gauging in three-dimensions for a massive self-dual vector multiplet. In ten-dimensions, we couple a single vector multiplet to another vector multiplet, where the latter gauges the scale covariance of the former. Due to scale covariance, the system does not have a lagrangian formulation, but has only a set of field equations, like Type IIB supergravity in ten-dimensions. As by-products, we construct similar models in six-dimensions with $N = (2,0)$ supersymmetry, and four-dimensions with $N = 1$ supersymmetry. We finally get a similar model with $N = 4$ supersymmetry in four-dimensions with consistent interactions that have never been known before. We expect a series of descendant theories in dimensions lower than ten by dimensional reductions. This result also indicates that similar mechanisms will work for other vector and scalar multiplets in space-time lower than ten-dimensions.

---

[1]E-Mail: hnishino@csulb.edu

[2]E-Mail: rajpoot@csulb.edu

## 1. Introduction

We have recently presented a model of gauged scale invariance for self-dual massive vector multiplet (VM) or scalar multiplet (SM) in three-dimensions (3D) [1]. In this formulation, we have basically two supermultiplets, *e.g.*, a SM and a GM. The former has a global nontrivial scaling properties that can be gauged by the latter. The scale covariance we introduced in [1] was different from the conventional dilatation [2]. One difference is that our scale transformation commutes with translation, while the conventional one does not [2]. Another difference is that we assign the same scaling weight for all the fields in a given supermultiplet, while in conformally supersymmetric models, the fermions and bosons differ their weights by 1/2 [3]. We have shown in [1] that supersymmetry and scale covariance are consistent with each other, both in component and superspace languages. We have also seen in [1] that such a system has no lagrangian, but has only a set of field equations. Moreover, the field equations for the GM can be free field equations, or can have nontrivial Dirac-Born-Infeld (DBI) type interactions [4], without upsetting the mutual consistency with the SM field equations.

Base on this development, the next natural question is whether such gaugings of scale covariance are universal in higher-dimensional globally supersymmetric theories. The most important system to study is $N=1$ globally supersymmetric VM in 10D, because any new theory established in 10D will generate similar descendant theories by simple dimensional reductions.

In this paper, we will present such a model in 10D. Namely, we show that we can gauge the scale covariance of a VM $(A_a, \lambda^\alpha)$ by an additional GM $(B_a, \chi^\alpha)$. We will formulate this model in terms of superspace, and investigate the consistency of field equations. The basic structure will turn out to be very similar to the model of gauged scale covariance in 3D [1]. This is a counter-example against the common wisdom that any consistent interactions with supersymmetry in 10D should be non-Abelian Yang-Mills theory [5], or DBI-type [4][6], related to superstring theory [5][7].

As by-products, we also present similar formulations in 6D with $N=(2,0)$ and 4D with $N=1$ supersymmetries. We finally perform a dimensional reduction of the 10D $N=1$ model into $N=4$ model in 4D with nontrivial interactions which has never been known before.

## 2. Superspace Formulation in 10D with $N=1$ Supersymmetry

We prepare basic relationships in superspace for our first model in 10D. We have two supermultiplets: the VM $(A_a, \lambda^\alpha)$ and the GM $(B_a, \chi^\alpha)$. In our 10D superspace notation, we use the indices $A \equiv (a,\alpha)$, $B \equiv (b,\beta)$, $\cdots$, with $a, b, \cdots = 0, 1, \cdots, 9$ for bosonic coordinates, and $\alpha, \beta, \cdots = 1, 2, \cdots, 16$ for chiral fermionic coordinates. Our metric in 10D is $(\eta_{ab}) = $ diag. $(-,+,\cdots,+)$ with Clifford algebra $\{\gamma_a, \gamma_b\} = +2\eta_{ab}$. In 10D, the charge conjugation matrix is anti-symmetric and chirality-flipping: $C_{\alpha\dot\beta} = -C_{\dot\beta\alpha}$ [8], so that the raising/lowering of spinor indices changes the chiralities, e.g., $\lambda^\alpha = C^{\alpha\dot\beta}\lambda_{\dot\beta}$. The GM is used to gauge the scale covariance of the VM. We formulate our model in terms of superspace [9] with global $N=1$ supersymmetry with the basic (anti)commutators

$$[\nabla_A, \nabla_B\} = T_{AB}{}^C \nabla_C - g G_{AB} \mathcal{S} \ , \tag{2.1}$$

where $g$ is a coupling constant, while the supercovariant derivative is defined by

$$\nabla_A \equiv E_A{}^M \partial_M - g B_A \mathcal{S} \equiv D_A - g B_A \mathcal{S} \ . \tag{2.2}$$

The $G_{AB}$ is the superfield strength of the potential superfield $B_A$, while $\mathcal{S}$ is the generator of scale transformation acting as

$$\mathcal{S} A_a = +A_a \ , \quad \mathcal{S} \lambda_\alpha = +\lambda_\alpha \ ,$$
$$\mathcal{S} G_{ab} = 0 \ , \quad \mathcal{S} \chi_\alpha = 0 \ . \tag{2.3}$$

Both VM and the GM has the superfield strengths:

$$F_{AB} \equiv \nabla_{[A} A_{B)} - T_{AB}{}^C A_C \ , \tag{2.4a}$$

$$G_{AB} \equiv \nabla_{[A} B_{B)} - T_{AB}{}^C B_C \ , \tag{2.4b}$$

satisfying the Bianchi identities (BIs)

$$\tfrac{1}{2}\nabla_{[A} F_{BC)} - \tfrac{1}{2} T_{[AB|}{}^D F_{D|C)} + \tfrac{1}{2} g G_{[AB} A_{C)} \equiv 0 \ , \tag{2.5a}$$

$$\tfrac{1}{2}\nabla_{[A} G_{BC)} - \tfrac{1}{2} T_{[AB|}{}^D G_{D|C)} \equiv 0 \ , \tag{2.5b}$$

$$\tfrac{1}{2}\nabla_{[A} T_{BC)}{}^D - \tfrac{1}{2} T_{[AB}{}^E T_{E|C)}{}^E - \tfrac{1}{4} R_{[AB|e}{}^f (\mathcal{M}_f{}^e)_{|C)}{}^D \equiv 0 \ . \tag{2.5c}$$

The potential superfield $A_A$ has its proper $U(1)$ gauge symmetry:

$$\delta_\Lambda A_A = \nabla_A \Lambda \equiv E_A{}^M \partial_M \Lambda - g B_A \Lambda \ . \tag{2.6}$$

where $\Lambda$ is a real scalar infinitesimal parameter superfield. Note that the $\nabla_A \Lambda$ contains the $B_A$-term, because even $\Lambda$ has the same scaling weight one as $A_A$. Due to the nontrivial coupling to the GM in (2.2), the superfield strength $F_{AB}$ is no longer invariant under (2.6), but instead transforming as

$$\delta_\Lambda F_{AB} = -g\Lambda G_{AB} \ . \tag{2.7}$$

The potential superfield $B_A$ gauges the scale covariance with the a real scalar infinitesimal parameter superfield $\Xi$ as

$$\delta_\Xi B_A = \nabla_A \Xi \ , \quad \delta_\Xi A_A = +g\Xi A_A \ , \quad \delta_\Xi F_{AB} = +g\Xi F_{AB} \ , \tag{2.8}$$

similarly to our 3D case [1].

As usual in superspace formulation [9], we need constraints which are listed as

$$T_{\alpha\beta}{}^c = +2(\gamma^c)_{\alpha\beta} \ , \quad F_{\alpha\beta} = G_{\alpha\beta} = 0 \ , \quad R_{ABc}{}^d = 0 \ , \tag{2.9a}$$

$$F_{\alpha b} = -(\gamma_b)_{\alpha\beta}\lambda^\beta \equiv +(\gamma_b \lambda)_\alpha \ , \quad G_{\alpha b} = -(\gamma_b)_{\alpha\beta}\chi^\beta \equiv +(\gamma_b \lambda)_\alpha \ , \tag{2.9b}$$

$$\nabla_\alpha \lambda^\beta = -\tfrac{1}{2}(\gamma^{cd})_\alpha{}^\beta F_{cd} + g\chi^\beta A_\alpha \ , \tag{2.9c}$$

$$\nabla_\alpha \chi^\beta = -\tfrac{1}{2}(\gamma^{cd})_\alpha{}^\beta G_{cd} \ , \tag{2.9d}$$

$$\nabla_\alpha F_{bc} = -(\gamma_{[b}\nabla_{c]}\lambda)_\alpha - g(\gamma_{[b|}\chi)_\alpha A_{|c]} - gG_{bc}A_\alpha \ , \tag{2.9e}$$

$$\nabla_\alpha G_{bc} = -(\gamma_{[b}\nabla_{c]}\chi)_\alpha \ . \tag{2.9f}$$

The important new feature here is the presence of the $g\chi A$-term in (2.9c) related to our gauged scale covariance. This term is required for the satisfaction of the $F$-BI at the engineering dimension $d=1$. These constraints are analogous to the 3D case [1].

There are some remarks about the presence of $A_c$ or $A_\alpha$ in the constraint (2.9c,e). The involvements of these bare potentials look unusual at first glance, but they are understood from the viewpoint of nontrivial transformation properties of $F_{AB}$ as in (2.7). In fact, (2.7) with (2.9b) gives us the transformation rule

$$\delta_\Lambda \lambda^\alpha = -g\Lambda\chi^\alpha \ , \tag{2.10}$$

which in turn explains the necessity of the terms with $A_c$ and $A_\alpha$. This is because for $\delta_\Lambda(\nabla_\alpha F_{bc})$ we get

$$\delta_\Lambda(\nabla_\alpha F_{bc}) = \nabla_\alpha(-gG_{bc}\Lambda) = +g\Lambda(\gamma_{[b}\nabla_{c]}\chi)_\alpha - gG_{bc}\nabla_\alpha\Lambda \ , \tag{2.11}$$

with the gradient term $\nabla_\alpha \Lambda$. On the other hand, if we take $\delta_\Lambda$ of (2.9e) using (2.10), we see how $\nabla_c \Lambda$ is cancelled by $\delta_\Lambda A_c$, while $\nabla_\alpha \Lambda$ is cancelled by $\delta_\Lambda A_\alpha$. These considerations justify the necessity of $A_c$ and $A_\alpha$-terms in (2.9c,e).

As has been already mentioned, our system does not have a lagrangian. This is because the scale covariance of the fermion $\lambda$ or the vector $B_a$ forbids the usual kinetic term of these fields. This is also related to the absence of gravity (zehnbein) in 10D that could be used to compensate such scaling at the lagrangian level. Therefore, the derivation of field equations for the VM and GM is imperative in our model.

Based on the preliminaries so far, we can derive the superfield equations from $d = 3/2$ BIs and higher. For example, the $\lambda$-field equation is obtained by the usual method of evaluating each side of the trivial identity $\{\nabla_\alpha, \nabla_\beta\}\lambda^\beta \equiv \nabla_{(\alpha}(\nabla_{\beta)}\lambda^\beta)$, and equate them. The $F$- or $G$-field equation is then obtained by applying spinorial derivatives on $\lambda$- or $\chi$-field equations. The field equations thus obtained are listed up as

$$(\slashed{\nabla}\lambda)_\alpha + g(\gamma^b \chi)_\alpha A_b \doteq 0 \ , \tag{2.12a}$$

$$\nabla_b F^{ba} - g(\overline{\lambda}\gamma^a \chi) - g G^{ab} A_b + \tfrac{1}{16}(\gamma^a \slashed{\nabla}\chi)^\beta A_\beta \doteq 0 \ , \tag{2.12b}$$

$$(\slashed{\nabla}\chi)_\alpha \doteq 0 \ , \tag{2.12c}$$

$$\nabla_b G^{ab} \doteq 0 \ , \tag{2.12d}$$

where the symbol $\doteq$ stands for a field equation.

The involvement of the bare potential $A_b$ in (2.12a) can be again understood by the peculiar transformation (2.10) of $\lambda$ under $\delta_\Lambda$. Similarly, the involvement of $A_b$ in (2.12b) is nothing bizarre, due to the transformation (2.7) of $F_{ab}$.

Even though the field equations for the GM are free, this situation is very similar to our recent results in 3D [1]. In 3D, we can further introduce some nontrivial interactions without upsetting the mutual consistency between the VM and GM. In 10D, however, there is some subtlety about this to be seen shortly.

The field equations (2.12) are of course scale covariant. For example, each bilinear interaction term has only the combination of a VM field and a GM field, because the kinetic term carries the unit scaling weight that should be the same for these interaction terms. Additionally, this also provides another explanation why there is no interaction terms between the VM and GM in the r.h.s. of (2.12c,d). Because any field in the VM carries a non-zero scaling weight, while the kinetic terms have zero scaling weight for these GM fields.

We can confirm the consistency of these field equations, *e.g.*, by taking the bosonic divergence of (2.12b) as the current conservation:

$$\nabla_a \Big[ \nabla_b F^{ba} - g(\overline{\lambda}\gamma^a\chi) - G^{ab}A_b + \tfrac{1}{16}g(\gamma^a\slashed{\nabla}\chi)^\beta A_\beta \Big]$$
$$= -g(\overline{\lambda}\slashed{\nabla}\chi) - g(\nabla_a G^{ab})A_b + g\overline{\chi}(\slashed{\nabla}\lambda + g\gamma^a\chi A_a) + \tfrac{1}{16}g\nabla_a\Big[(\gamma^a\slashed{\nabla}\chi)^\beta A_\beta\Big]$$
$$\doteq 0 \ . \tag{2.13}$$

This vanishes by the use of field equations in (2.12), showing the mutual consistency. Additionally, the structure of vanishing of each term in (2.13) tells us why the GM field equations are to be free. In contrast to the 3D case [1] where the VM had no proper gauge covariance, we have now the conservation of current for the proper $U(1)$ covariance (2.6) for the VM. The conservation of its current seems to require that the GM field equations are to be free.

Another nontrivial confirmation is the fermionic derivative applied on (2.12b):

$$\nabla_\alpha \Big[ \nabla_b F^{ba} - g(\overline{\lambda}\gamma^a\chi) - gG^{ab}A_b + \tfrac{1}{16}(\gamma^a\slashed{\nabla}\chi)^\beta A_\beta \Big]$$
$$= -\nabla^a(\slashed{\nabla}\lambda + g\gamma^b\chi A_b)_\alpha + \Big[\gamma^a\slashed{\nabla}(\slashed{\nabla}\lambda + g\gamma^b\chi A_b)\Big]$$
$$+ g(\gamma^a\gamma^b\slashed{\nabla}\chi)_\alpha A_b - g(\slashed{\nabla}\chi)_\alpha A_a - g(\nabla_b G^b{}_a)A_\alpha + \tfrac{1}{16}g\nabla_\alpha\Big[(\gamma^a\slashed{\nabla}\chi)^\beta A_\beta\Big]$$
$$\doteq 0 \ . \tag{2.14}$$

It is clear that each term in (2.14) vanishes by the use of the field equations (2.12), in particular, the free GM field equations (2.12c,d).

## 3. Gauged Scale Covariance in 6D with $N = (2,0)$ Supersymmetry

Once we have understood the mechanism of gauged scale covariance in 10D, we can try similar formulations in lower-dimensions. The first good example is 6D with $N = (2,0)$ which is the minimal number of supersymmetries in there. The reason we choose 6D is that an $N = (2,0)$ VM in 6D has the field content $(A_a, \lambda_{\alpha A})$ with no scalars which might complicate the computation. For example, VMs have scalar(s) in space-time dimensions $9 \geq D \geq 7$ [10]. In other words, 6D is the next space-time dimension lower than 10D where there is no scalar in a VM with simple supersymmetry.

As described above, the field content of a VM is $(A_a, \chi^{\alpha A})$, where the index $A = 1, 2$ is for the **2** of $Sp(1)$ [11], while the superscript indices $\alpha, \beta, \cdots = 1, 2, \cdots, 8$ are for the spinorial index with the positive chirality. In other words, the VM has a pair of Majorana-Weyl spinors with positive chirality forming the **2** of $Sp(1)$. The GM for gauging has the field content $(B_a, \chi^{\alpha A})$, where again $\chi^{\alpha A}$ has the positive chirality. As universal in this paper, all the fields in the VM have scaling weight $+1$, while those in the GM have zero scaling weight.

In 6D with the metric $(\eta_{ab}) = \text{diag.}(-,+,\cdots,+)$, we have the anti-symmetric gamma matrix $(\gamma^c)_{\alpha\beta} = -(\gamma^c)_{\beta\alpha}$ [8], while the charge conjugation matrix is also anti-symmetric $C_{\dot\alpha\dot\beta} = -C_{\dot\beta\dot\alpha}$ where the dotted index $\dot\beta$ stands for the negative chirality. Accordingly, the raising/lowering of spinor indices changes their chiralities as in 10D: e.g., $\lambda^{\alpha A} = C^{\alpha\dot\beta}\lambda_{\dot\beta}{}^A$. The raising/lowering of $Sp(1)$ indices are done by the $Sp(1)$ metric $\epsilon_{AB}$,[3] such as $\lambda^{\alpha A} \equiv \epsilon^{AB}\lambda^\alpha{}_B$, $(\overline\chi\gamma^a\lambda) \equiv -\chi^{\alpha A}(\gamma^a)_{\alpha\beta}\lambda^\beta{}_A \equiv -\epsilon^{AB}\chi^\alpha{}_B(\gamma^a)_{\alpha\beta}\lambda^\beta{}_A$. We sometimes use the underlined indices $\underline\alpha, \underline\beta, \cdots \equiv (\alpha,A), (\beta,B),\cdots$ for the combination of the chiral indices $\alpha, \beta, \cdots$ and the $Sp(1)$ indices $A, B, \cdots$. For example, $(\gamma^c)_{\underline{\alpha\beta}} \equiv (\gamma^c)_{\alpha\beta}\epsilon_{AB}$, or $C_{\underline{\alpha\beta}} \equiv C_{\dot\alpha\dot\beta}\epsilon_{AB}$ is the charge conjugation matrix with the $Sp(1)$ indices combined.

Our basic BIs in 6D are formally the same as (2.5). Even the involvement of the $gGA$-term in the $F$-BIs is the same. The basic constraints in $N=(2,0)$ superspace for 6D are

$$T_{\underline{\alpha\beta}}{}^c = +2(\gamma^c)_{\underline{\alpha\beta}} = +2(\gamma^c)_{\alpha\beta}\epsilon_{AB} , \quad F_{\underline{\alpha\beta}} = G_{\underline{\alpha\beta}} = 0 , \quad R_{ABc}{}^d = 0 , \quad (3.1\text{a})$$

$$F_{\underline\alpha b} = +(\gamma_b)_{\underline{\alpha\beta}}\lambda^{\underline\beta} \equiv +(\gamma_b\lambda)_{\underline\alpha} , \quad G_{\underline\alpha b} = +(\gamma_b)_{\underline{\alpha\beta}}\chi^{\underline\beta} \equiv +(\gamma_b\chi)_{\underline\alpha} , \quad (3.1\text{b})$$

$$\nabla_{\underline\alpha}\lambda^{\underline\beta} = -\tfrac{1}{2}(\gamma^{cd})_{\underline\alpha}{}^{\underline\beta}F_{cd} + g\chi^{\underline\beta}A_{\underline\alpha} \equiv -\tfrac{1}{2}(\gamma^{cd})_\alpha{}^\beta\delta_A{}^B F_{cd} + g\chi^{\underline\beta}A_{\underline\alpha} , \quad (3.1\text{c})$$

$$\nabla_{\underline\alpha}\chi^{\underline\beta} = -\tfrac{1}{2}(\gamma^{cd})_{\underline\alpha}{}^{\underline\beta}G_{cd} \equiv -\tfrac{1}{2}(\gamma^{cd})_\alpha{}^\beta\delta_A{}^B G_{cd} , \quad (3.1\text{d})$$

$$\nabla_{\underline\alpha}F_{bc} = -(\gamma_{[b}\nabla_{c]}\lambda)_{\underline\alpha} - g(\gamma_{[b|}\chi)_{\underline\alpha}A_{|c]} - gG_{bc}A_{\underline\alpha} , \quad (3.1\text{e})$$

$$\nabla_{\underline\alpha}G_{bc} = -(\gamma_{[b}\nabla_{c]}\chi)_{\underline\alpha} . \quad (3.1\text{f})$$

Note the subtle difference in signatures for spinorial multiplications. For example in (3.1b), we have a positive sign for $(\gamma_b)_{\underline{\alpha\beta}}\lambda^{\underline\beta}$, because in terms of *underlined* fermionic indices, the charge conjugation matrix is symmetric: $C_{\underline{\alpha\beta}} = C_{\underline{\beta\alpha}}$.

The field equations for our $N=(2,0)$ system in 6D are

$$(\slashed\nabla\lambda)_{\underline\alpha} + g(\gamma^b\chi)_{\underline\alpha}A_b \doteq 0 , \quad (3.2\text{a})$$

$$\nabla_b F^{ba} - g(\overline\lambda\gamma^a\chi) - gG^{ab}A_b + \tfrac{1}{8}(\gamma^a\slashed\nabla\chi)^{\underline\beta}A_{\underline\beta} \doteq 0 , \quad (3.2\text{b})$$

$$(\slashed\nabla\chi)_{\underline\alpha} \doteq 0 , \quad (3.2\text{c})$$

$$\nabla_b G^{ab} \doteq 0 . \quad (3.2\text{d})$$

---
[3]We believe that the indices $A, B, \cdots$ for the **2** of $Sp(1)$ are not confusing with the superspace indices, as long as we are careful about the context.

These are parallel to (2.12), showing the similar structures in the gauging. The only differences is the coefficient $1/8$ in (3.2b) instead of $1/16$, caused by the trace $(\gamma^b\gamma^a\gamma^{cd})^\gamma{}_{\underline{\gamma}}\nabla_b F_{cd}$, *etc.*, depending on the range of spinorial indices.

We see not only the scaling weights but also chiralities are consistent in these field equations. At first glance, the fact that even the signatures of all the terms are the same as the 10D case is amazing at first glance. However, this is understandable from the viewpoint of simple dimensional reduction from 10D. As a matter of fact, such parallel structures are expected in dimensional reduction even in superspace [12].

## 4. Gauged Scale Covariance in 4D with $N=1$ Supersymmetry

We next study a similar gauging in 4D. Going down from 6D, we find 4D is the next dimensions, where a VM has no scalar. In fact, in 5D there is a scalar field needed for $N=2$ VM with $4+4$ degrees of freedom.

As is well-known in 4D, a VM with $N=1$ supersymmetry has the field content $(A_a, \lambda_\alpha)$ with a Majorana spinor $\lambda_\alpha$. Our GM has the field content $(B_a, \chi_\alpha)$ again with a Majorana spinor $\chi_\alpha$. In this section, we use the spinorial indices $\alpha, \beta, \cdots = 1, \cdots, 4$ in our 4D space time with $(\eta_{ab}) = \text{diag.}\,(-,+,+,+)$. As before, all the fields in the VM have the unit scaling weight, while those in the GM have zero scaling weight.

Our basic constraints are very similar to the 10D case (2.9):

$$T_{\alpha\beta}{}^c = +2(\gamma^c)_{\alpha\beta} \ , \quad F_{\alpha\beta} = G_{\alpha\beta} = 0 \ , \quad R_{ABc}{}^d = 0 \ , \tag{4.1a}$$

$$F_{\alpha b} = +(\gamma_b)_\alpha{}^\beta \lambda_\beta \equiv +(\gamma_b \lambda)_\alpha \ , \quad G_{\alpha b} = +(\gamma_b)_\alpha{}^\beta \chi_\beta \equiv +(\gamma_b \chi)_\alpha \ , \tag{4.1b}$$

$$\nabla_\alpha \lambda_\beta = -\tfrac{1}{2}(\gamma^{cd})_{\alpha\beta} F_{cd} + g\chi_\beta A_\alpha \ , \tag{4.1c}$$

$$\nabla_\alpha \chi_\beta = -\tfrac{1}{2}(\gamma^{cd})_{\alpha\beta} G_{cd} \ , \tag{4.1d}$$

$$\nabla_\alpha F_{bc} = -(\gamma_{[b}\nabla_{c]}\lambda)_\alpha - g(\gamma_{[b|}\chi)_\alpha A_{|c]} - g G_{bc} A_\alpha \ , \tag{4.1e}$$

$$\nabla_\alpha G_{bc} = -(\gamma_{[b}\nabla_{c]}\chi)_\alpha \ . \tag{4.1f}$$

The field equations for our $N=1$ system in 4D are

$$(\slashed{\nabla}\lambda)_\alpha + g(\gamma^b \chi)_\alpha A_b \doteq 0 \ , \tag{4.2a}$$

$$\nabla_b F^{ba} - g(\overline{\lambda}\gamma^a \chi) - g G^{ab} A_b + \tfrac{1}{4}(\gamma^a \slashed{\nabla}\chi)^\beta A_\beta \doteq 0 \ , \tag{4.2b}$$

$$(\slashed{\nabla}\chi)_\alpha \doteq 0 \ , \tag{4.2c}$$

$$\nabla_b G^{ab} \doteq 0 \ . \tag{4.2d}$$

Here we have the coefficient $1/4$ in (4.2b) instead of $1/16$ in the 10D case.

## 5. Gauged Scale Covariance in 4D with $N = 4$ Supersymmetry

We finally present a nontrivial model of gauged scale covariance with $N = 4$ supersymmetry in 4D. This will give us new consistent interactions in 4D with $N = 4$ supersymmetry that have not been presented before, ever since the first $N = 4$ non-Abelian model [5]. This can be derived by a dimensional reduction of our model in 10D. We have two multiplets as before, but now with richer field contents: VM $(A_a, \lambda_{\alpha(i)}, A_i, \widetilde{A}_i)$ and GM $(B_a, \chi_{\alpha(i)}, B_i, \widetilde{B}_i)$, where $a, b, \cdots = 0, 1, 2, 3$ are for the 4D bosonic coordinates, $\alpha, \beta, \cdots = 1, 2, 3, 4$ are for fermionic coordinates. As is well-known, both of these multiplets have $8+8$ physical degrees of freedom. We use parentheses for the $N = 4$ indices $(i), (j), \cdots = (1), (2), (3), (4)$ distinct from $i, j, k, \cdots = 1, 2, 3$. The *tilded* spin-less fields $\widetilde{A}_i$ and $\widetilde{B}_i$ are pseudo-scalars. All the fields in the VM have the unit scaling weight, while those in the GM have zero scaling weight. Accordingly, our $N = 4$ superspace has the indices $A \equiv (a, \underline{\alpha})$, $B \equiv (b, \underline{\beta}), \cdots$ where the fermionic indices are $\underline{\alpha} \equiv \alpha(i), \underline{\beta} \equiv \beta(j), \cdots$. For example, our superderivatives are

$$\nabla_A \equiv E_A{}^M \partial_M - g B_A \mathcal{S} \equiv D_A - g B_A \mathcal{S} \ , \tag{5.1}$$

where the fermionic ones are $\nabla_{\underline{\alpha}} \equiv \nabla_{\alpha(i)}$ $\big((i) = (1), (2), (3), (4)\big)$ corresponding to $N = 4$ supersymmetry.

For the notation for our dimensional reductions, we always use the *hat*-symbol on the fields and indices in 10D to distinguish them from 4D ones, as usual in component field dimensional reduction [13]. Our dimensional reduction in superspace is also similar to that has been performed in [12]. The basic dimensional reduction rules are summarized as follows: First of all, the vector fields are reduced as $(\widehat{A}_{\hat{a}}) = (\widehat{A}_a, \widehat{A}_{3+i}, \widehat{A}_{6+i}) \equiv (A_a, A_i, \widetilde{A}_i)$, $(\widehat{B}_{\hat{a}}) = (\widehat{B}_a, \widehat{B}_{3+i}, \widehat{B}_{6+i}) \equiv (B_a, B_i, \widetilde{B}_i)$ with $i = 1, 2, 3$. Accordingly, we have

$$\widehat{F}_{\hat{a}\hat{b}} = \begin{cases} \widehat{F}_{ab} = F_{ab} \ , \\ \widehat{F}_{a,3+i} = \nabla_a A_i + g B_i A_a \ , \\ \widehat{F}_{a,6+i} = \nabla_a \widetilde{A}_i + g \widetilde{B}_i A_a \ , \\ \widehat{F}_{3+i,3+j} = -g(B_i A_j - B_j A_i) \ , \\ \widehat{F}_{3+i,6+j} = -g(B_i \widetilde{A}_j - \widetilde{B}_j A_i) \ , \\ \widehat{F}_{6+i,6+j} = -g(\widetilde{B}_i \widetilde{A}_j - \widetilde{B}_j \widetilde{A}_i) \ , \end{cases} \qquad \widehat{G}_{\hat{a}\hat{b}} = \begin{cases} \widehat{G}_{ab} = G_{ab} \ , \\ \widehat{G}_{a,3+i} = \nabla_a B_i \ , \\ \widehat{G}_{a,6+i} = \nabla_a \widetilde{B}_i \ , \\ \text{Otherwise } 0 \ , \end{cases} \tag{5.2}$$

for $i = 1, 2, 3$. As for $\gamma$-matrices, we follow ref. [5] as

$$\widehat{\gamma}_{\widehat{a}} = \begin{cases} \widehat{\gamma}_a = \sigma_3 \otimes I_4 \otimes \gamma_a \ , \\ \widehat{\gamma}_{3+i} = \sigma_2 \otimes \beta_3 \alpha_i \otimes I_4 \ , \\ \widehat{\gamma}_{6+i'} = I_2 \otimes \beta_{i'} \otimes \gamma_5 \quad (i' = 1,\, 2) \ , \\ \widehat{\gamma}_9 = \sigma_3 \otimes \beta_3 \otimes \gamma_5 \ , \end{cases} \tag{5.3}$$

in the $32 \times 32$ representation in 10D with $n \times n$ unit matrix $I_n$. Accordingly, we have also $\widehat{\gamma}_{11} = -\sigma_1 \otimes \beta_3 \otimes I$. The $\sigma_i$ are the usual $2 \times 2$ Pauli matrices, and $\alpha$'s and $\beta$'s are $4 \times 4$ matrices forming the generators of $SO(3) \times SO(3)$, satisfying [5]

$$\alpha_i \alpha_j = \delta_{ij} + i\epsilon_{ijk}\alpha_k \ , \quad \beta_i \beta_j = \delta_{ij} + i\epsilon_{ijk}\beta_k \ , \quad [\alpha_i, \beta_j] = 0 \ . \tag{5.4}$$

Both $\widehat{\lambda}$ and $\widehat{\chi}$-fields have the positive chirality in 10D: $\widehat{\gamma}_9 \widehat{\lambda} = +\widehat{\lambda}$, $\widehat{\gamma}_9 \widehat{\chi} = +\widehat{\chi}$, so that we have their reduction rule

$$\widehat{\lambda} = \begin{pmatrix} \lambda \\ -\beta_3 \lambda \end{pmatrix} \ , \quad \widehat{\chi} = \begin{pmatrix} \chi \\ -\beta_3 \chi \end{pmatrix} \ , \tag{5.5}$$

where $\lambda$ and $\chi$ have implicit indices $\alpha(i)$. In other words, the original Majorana-Weyl spinors $\widehat{\lambda}^{\widehat{\alpha}}$ and $\widehat{\chi}^{\widehat{\alpha}}$ in 10D are decomposed into 4 copies of 4-component Majorana spinors $\lambda_{\alpha(i)}$ and $\chi_{\alpha(i)}$ in 4D.[4]

Our constraints in 4D superspace are dictated as

$$T_{\underline{\alpha}\underline{\beta}}{}^c = +2(\gamma^c)_{\alpha\beta}\, \delta_{(i)(j)} \equiv +2(\gamma^c)_{\underline{\alpha}\underline{\beta}} \ , \quad F_{\underline{\alpha}\underline{\beta}} = G_{\underline{\alpha}\underline{\beta}} = 0 \ , \quad R_{AB c}{}^d = 0 \ , \tag{5.6a}$$

$$F_{\underline{\alpha} b} = +(\gamma_b)_{\underline{\alpha}}{}^{\underline{\beta}} \lambda_{\underline{\beta}} \equiv +(\gamma_b \lambda)_{\underline{\alpha}} \ , \quad G_{\underline{\alpha} b} = +(\gamma_b)_{\underline{\alpha}}{}^{\underline{\beta}} \chi_{\underline{\beta}} \equiv +(\gamma_b \chi)_{\underline{\alpha}} \ , \tag{5.6b}$$

$$\nabla_{\underline{\alpha}} \lambda_{\underline{\beta}} = -\tfrac{1}{2}(\gamma^{cd})_{\underline{\alpha}\underline{\beta}} F_{cd} + g\chi_{\underline{\beta}} A_{\underline{\alpha}} + i(\alpha_i \gamma^a)_{\underline{\alpha}\underline{\beta}}(\nabla_a A_i + gB_i A_a)$$
$$+ (\beta_i \gamma_5 \gamma^a)_{\underline{\alpha}\underline{\beta}}(\nabla_a \widetilde{A}_i + g\widetilde{B}_i A_a) + ig(\alpha_i \beta_j \gamma_5)_{\underline{\alpha}\underline{\beta}}(B_i \widetilde{A}_j - \widetilde{B}_j A_i)$$
$$+ ig\epsilon_{ijk}(\alpha_k)_{\underline{\alpha}\underline{\beta}} B_i A_j + ig\epsilon_{ijk}(\beta_k)_{\underline{\alpha}\underline{\beta}} \widetilde{B}_i \widetilde{A}_j \ , \tag{5.6c}$$

$$\nabla_{\underline{\alpha}} \chi_{\underline{\beta}} = -\tfrac{1}{2}(\gamma^{cd})_{\alpha\beta}\delta_{(i)(j)} G_{cd} + i(\alpha_i \gamma^a)_{\underline{\alpha}\underline{\beta}} \nabla_a B_i + (\beta_i \gamma_5 \gamma^a)_{\underline{\alpha}\underline{\beta}} \nabla_a \widetilde{B}_i \ , \tag{5.6d}$$

$$\nabla_{\underline{\alpha}} A_i = +i(\alpha_i)_{(i)(j)} \lambda_{\alpha(j)} \equiv +i(\alpha_i \lambda)_{\underline{\alpha}} \ , \tag{5.6e}$$

$$\nabla_{\underline{\alpha}} \widetilde{A}_i = +(\beta_i)_{(i)(j)} (\gamma_5 \lambda)_{\alpha(j)} \equiv +(\beta_i \gamma_5 \lambda)_{\underline{\alpha}} \ , \tag{5.6f}$$

$$\nabla_{\underline{\alpha}} B_i = +i(\alpha_i)_{(i)(j)} \chi_{\alpha(j)} \equiv +i(\alpha_i \chi)_{\underline{\alpha}} \ , \tag{5.6g}$$

$$\nabla_{\underline{\alpha}} \widetilde{B}_i = +(\beta_i)_{(i)(j)} (\gamma_5 \chi)_{\alpha(j)} \equiv +(\beta_i \gamma_5 \chi)_{\underline{\alpha}} \ . \tag{5.6h}$$

---

[4]In the expression (5.5), both $\widehat{\lambda}$ and $\widehat{\chi}$ have 32 components, corresponding to the $32 \times 32$ matrix $\widehat{\gamma}_9$, but they have effectively 16 components, due to their chiralities [5].

Here we use the simplified expressions, such as, $(\alpha_i)_{\underline{\alpha\beta}} \equiv C_{\alpha\beta}(\alpha_i)_{(i)(j)}$, etc. These forms reflect the original structures in 10D, but also have the new effects of our dimensional reduction.

Our field equations are listed as

$$\slashed{\nabla}\lambda - ig\alpha_i\lambda B_i - g\beta_i\gamma_5\lambda\widetilde{B}_i + g\gamma^a\chi A_a + ig\alpha_i\chi A_i + g\beta_i\gamma_5\chi\widetilde{A}_i \doteq 0 \ , \tag{5.7a}$$

$$\nabla_b F^{ba} + gB_i(\nabla^a A_i + gB_i A^a) + g\widetilde{B}_i(\nabla^a \widetilde{A}_i + g\widetilde{B}_i A^a)$$
$$- g(\overline{\lambda}\gamma^a\chi) - gG^{ab}A_b - gA_i\nabla^a B_i - g\widetilde{A}_i\nabla^a\widetilde{B}_i \doteq 0 \ , \tag{5.7b}$$

$$\nabla_a(\nabla^a A_i + gB_i A^a) - ig(\overline{\lambda}\alpha_i\chi) + gA^a\nabla_a B_i$$
$$+ g^2 B_j(B_j A_i - B_i A_j) - g^2 \widetilde{B}_j(B_i\widetilde{A}_j - \widetilde{B}_j A_i) \doteq 0 \ , \tag{5.7c}$$

$$\nabla_a(\nabla^a \widetilde{A}_i + g\widetilde{B}_i A^a) - g(\overline{\lambda}\beta_i\gamma_5\chi) + gA^a\nabla_a\widetilde{B}_i$$
$$+ g^2 B_j(B_j\widetilde{A}_i - \widetilde{B}_i A_j) - g^2\widetilde{B}_j(\widetilde{B}_i\widetilde{A}_j - \widetilde{B}_j\widetilde{A}_i) \doteq 0 \ , \tag{5.7d}$$

$$\slashed{\nabla}\chi \doteq 0 \ , \tag{5.7e}$$

$$\nabla_b G^{ab} \doteq 0 \ , \tag{5.7f}$$

$$\nabla_a^2 B_i \doteq 0 \ , \quad \nabla_a^2 \widetilde{B}_i \doteq 0 \ . \tag{5.7g}$$

Here we have omitted the fermionic indices for fermionic field equations.

The particular combinations $\nabla_a A_i + gB_i A_a$ and $\nabla_a \widetilde{A}_i + g\widetilde{B}_i A_a$ in (5.7b,c,d) are important. Even though the involvement of the bare potential $A_a$ seems unusual, it can be understood as covariance under the $U(1)$ transformation $\delta_\Lambda$. In fact, $A_i$ and $\widetilde{A}_i$ transform nontrivially under $\delta_\Lambda$, as the original 10D rule shows:

$$\delta_\Lambda A_i = -g\Lambda B_i \ , \quad \delta_\Lambda \widetilde{A}_i = -g\Lambda \widetilde{B}_i \ , \tag{5.8}$$

so that we have

$$\delta_\Lambda(\nabla_a A_i + gB_i A_a) = -g\Lambda\nabla_a B_i \ , \quad \delta_\Lambda(\nabla_a\widetilde{A}_i + g\widetilde{B}_i A_a) = -g\Lambda\nabla_a\widetilde{B}_i \ , \tag{5.9}$$

with the factor $\nabla_a\Lambda$ cancelled, as desired. This is also understandable that these particular combinations correspond to $\widehat{F}_{a,3+i}$ and $\widehat{F}_{a,6+i}$ in (5.2) transforming as (2.7). By the same token, we see the necessity of the $A^a\nabla_a B_i$ and $A^a\nabla_a\widetilde{B}_i$-terms in (5.7c,d).

The validity of our field equations can be confirmed by taking a fermionic derivative $\nabla_{\underline{\alpha}}$ on the $\lambda$ or $\chi$-field equations. They produce terms that vanish by the use of other bosonic field equations, as usual in supersymmetric models. In this process, useful identities are

$$\delta_{(i)(j)}\delta_{(k)(\ell)} - \delta_{(k)(j)}\delta_{(i)(\ell)} \equiv -\tfrac{1}{2}(\alpha_i)_{(i)(k)}(\alpha_i)_{(j)(\ell)} - \tfrac{1}{2}(\beta_i)_{(i)(k)}(\beta_i)_{(j)(\ell)} \ , \quad (5.10\text{a})$$

$$\left[ (\alpha_i)_{(k)(i)}(\alpha_i)_{(j)(\ell)} - (\beta_i)_{(k)(i)}(\beta_i)_{(j)(\ell)} \right] + {}_{(i)\leftrightarrow(j)} \equiv 0 \ . \quad (5.10\text{b})$$

Needless to say, our previous $N=1$ model in 4D is obtained by a consistent truncation of $A_i = \widetilde{A}_i = B_i = \widetilde{B}_i = 0$, with reducing the components of $\lambda$ and $\chi$ from 16 to 4, by deleting the index $(i)$ on them.

The important point here is that the gauged scale covariance is compatible not only with $N=1$ simple supersymmetry, but also with $N=4$ extended supersymmetry. Since our $N=1$ model in 10D generates all the maximally extended global supersymmetries in lower-dimensions, this feature seems universal in diverse dimensions.

Note that our $N=4$ system has the dimensionless coupling $g$, similar to the conventional non-Abelian $N=4$ model [5]. This promotes our model not only to a renormalizable theory, but also to a plausible ultraviolet *finite* theory, just as the conventional $N=4$ models [14].

## 6. Concluding Remarks

In this paper we have presented a very peculiar model of gauging scale covariance with $N=1$ supersymmetry in 10D. We have seen that all the couplings are consistent with each other, even though the field equations for the GM can be free. This situation is very similar to our recent result in 3D [1].

As by-products, we have also constructed similar models of gauging scale covariances both in 6D and 4D respectively with $N=(2,0)$ and $N=1$ supersymmetries. All the relevant multiplets do not have scalar fields, so that the treatments in these dimensions are relatively easy. We have also performed dimensional reduction of our 10D model into $N=4$ model in 4D which has entirely new nontrivial couplings as a globally supersymmetric model in 4D.

We stress the crucial point that not only $N=1$ simple supersymmetry, but also $N=4$ extended supersymmetry is shown to be compatible with gauged scale covariance in 4D. In particular, we have seen in (5.8) that some scalar fields in the $N=4$ model are transforming nontrivially under the proper $U(1)$ symmetry of the VM, playing an important role. The compatibility of gauged scale covariance with maximally-extended global supersymmetries seems universal also in higher-dimensions, because our 10D model generates all such maximally extended models in space-time $D \leq 9$.

The success of our formulations is very encouraging from an additional viewpoint. Namely, other similar models with gauged scale covariance may be constructed in lower-dimensions which are *not* necessarily related to our 10D model by simple dimensional reductions. In fact, our $N=4$ case in 4D indicates that we can develop gauged scale covariance not only for VMs but also for scalar multiplets which exists in $D \leq 6$ [10]. In fact, our results in 3D [1] form a subset of such applications.

Our $N=4$ model in 4D is an interesting model, because it is most likely ultraviolet finite as a common feature of $N=4$ theories, like non-Abelian $N=4$ supersymmetric models [5][14]. Interestingly, we may well have an extra finite theory which is entirely new and different from the conventional non-Abelian $N=4$ supersymmetric model [5].

The result in this paper is very peculiar. Because the common wisdom in the past has been that no nontrivial interactions exist with $N=1$ global supersymmetry in 10D other than Yang-Mills type [5] or DBI-type [6]. This is because the possible consistent interactions among VMs are so tight that we can not easily modify their interactions. Or at least, any interaction is supposed to be related to superstring theory [7]. The model we presented in this paper has provided a counter-example against such common wisdom, namely, we have nontrivial interactions that have not been known before, and it does not yet have to be related to superstring theory [7]. We believe that our result in this paper provides a completely new angle to study general supersymmetric theories in diverse dimensions.

We are grateful to W. Siegel for important discussions and reading the draft. This work is supported in part by NSF Grant # 0308246.